\documentclass[prl,preprint]{revtex4}
\usepackage[utf8]{inputenc}

\usepackage{epsfig}
\usepackage{amsmath,multirow,amssymb,graphicx}
\usepackage{graphics}
\usepackage{graphicx}
\usepackage{amsmath}
\usepackage{pstricks}
\usepackage{subfigure}
\usepackage{epstopdf}

\begin{document}

\title{Active Two-dimensional Steering of Radiation from a Nano-Aperture}
\author{Laura~S.~Dreissen,$^1$ Hugo~F.~Schouten,$^1$ Wim~Ubachs,$^{1,2}$ 
Shreyas~B.~Raghunathan,$^3$ and Taco~D.~Visser$^{1,4,*}$ }

\affiliation{ \vspace*{0.5cm}
$^1$Department~of Physics and Astronomy, LaserLaB, \\Vrije Universiteit\\
De Boelelaan 1081, 1081 HV Amsterdam, The Netherlands\\
$^2$Advanced Research Center for Nanolithography, Science Park 110, \\1098 XG Amsterdam, The Netherlands\\
$^3$ASML, De Run  6501, \\
5504 DR Veldhoven, 
The Netherlands\\
$^4$Department~of Physics and Astronomy,~University of Rochester 
\\Rochester, NY 14627, USA\\
\\
$^*$Corresponding author: t.d.visser@vu.nl}

\begin{abstract}

We experimentally demonstrate control over the direction of radiation of a beam  
that passes through a square nano-aperture in a metal film. 
The ratio of the aperture size and the wavelength is such that only three guided modes, each with different 
spatial symmetries, can be excited. 
Using a spatial light modulator,  the superposition of the three modes can be altered, thus allowing for
a controlled variation of the radiation pattern that emanates from the nano-aperture. 
Robust and stable steering of $9.5^\circ$ in two orthogonal directions was achieved.

\vspace{0.5cm}

\end{abstract}

\maketitle
Since the seminal work by Bethe$^1$, many studies have been devoted to understanding the physical mechanisms that underly the transmission of radiation 
through nano-apertures$^{2-11}$, a topic of both fundamental and technical interest. 
Several methods have been proposed to dynamically 
control  the total transmission, in order to achieve all-optical switching$^{12-14}$. 
A useful next step would be to actively steer the radiation, something of great importance 
for applications such as selective probing of nano-samples, and all-optical circuits for telecommunication. 
Directional  transmission has been achieved, for example, by using a single sub-wavelength slit surrounded by surface corrugations or 
grooves$^{15-16}$, and by varying the refractive index of neighboring sub-wavelength 
slits$^{17}$. However, all these approaches lead to a {\em static} asymmetry of the radiated field. 
What has not been achieved so far, is the possibility to {\em dynamically} steer the transmitted field in two 
orthogonal directions.

Here we report an experiment in which such steering is clearly demonstrated. 
Central to our approach is the selective excitation of guided modes in a wavelength-sized aperture 
in a metal film. We previously used a similar technique 
to control the direction in which surface plasmons polaritons are launched$^{18}$, and to obtain one-dimensional beam steering
from a narrow slit$^{19}$. A non-trivial extension to two-dimensional steering has now 
been realized by 
using a square aperture illuminated by light with a wavelength such that only three guided modes can be excited. 
Using a spatial light modulator the phase of each of the  three modes is controlled. 
The transmitted field, which is a coherent superposition of these modes, can thus be altered, 
leading to a change in the directionality of the emanating field.

The radiation from an aperture can be understood by analyzing the guided (i.e., non-evanescent)  
modes that it can sustain. Consider a square hole with sides $a$, in a perfect conductor, such that 
\begin{align}
\lambda < a <  \frac{\sqrt{5}}{2}  \lambda , 
\label{size}
\end{align}
where $\lambda$ denotes the free-space wavelength. If the incident field of frequency $\omega$ is $x$-polarized, 
only three modes can be excited$^{20}$, namely 
${\rm TE}_{01}$, ${\rm TE}_{02}$, and a hybrid ${\rm TE}_{11}/{\rm TM}_{11}$ mode, 
the latter being such that $E_y = 0$. 
The $x$ component of the electric field of the $mn$-mode is given by the expression 
\begin{align}
E_{x;mn} ({\bf r},t)= A_{mn} \cos \left ( \frac{m \pi x}{a} \right ) \sin \left ( \frac{n \pi y}{a} \right ) 
\exp[{\rm i}(k_z z - \omega t)], 
\label{modes}
\end{align}
where ${\bf r} =(x,y,z)$ is a point in space, $t$ is a moment in time, $A_{mn}$ is an amplitude, and  $k_z$ denotes the effective longitudinal wavenumber of the mode. 
In Eq.~(\ref{modes}) the origin of the coordinates is taken at the bottom left corner of the aperture. 
The spatial symmetry properties of each mode with 
respect to the center of the aperture are listed in Table 1. 
It is seen that every mode displays a unique combination of symmetries in the $x$ and $y$ direction. 
As will be shown shortly, by considering modes with specific spatial symmetries in two dimensions, rather than one (as in$^{18,19}$), we 
can achieve 
active radiation steering in both the $x$ and $y$ direction. 

For our method it is essential that the aperture is square-shaped with sides given by Eq.~(\ref{size}). 
Only then does a degenerate, hybrid ${\rm TE}_{11}/{\rm TM}_{11}$ mode exist, and only then 
does the aperture allow exactly three guided modes with the right spatial symmetries. 
It is unclear if the same goal of selective excitation of non-evanescent modes with a prescribed 
symmetry could be achieved with a circular aperture.

\vspace*{0.7cm}
\begin{tabular}{lcc}
\multicolumn{3}{c}{Table 1: Symmetries of $E_x$ of guided modes 
} \\
\hline
mode    & $x$ direction & $y$ direction  \\
\hline
${\rm TE}_{01}$      & even    & even      \\
${\rm TE}_{02}$      & even    & odd      \\
${\rm TE_{11}/TM}_{11}$       &odd     & even      \\
\hline
\label{tabel}
\end{tabular}

The radiated electric field due to each mode can be calculated using the far-field diffraction formula$^{20}$ 
\begin{eqnarray}
{\bf E}_{mn}({\bf r})&= &\frac{{\rm i} e^{{\rm i} kr}}{2 \pi r} {\bf k} \times 
 \int_{-a/2}^{a/2} \int_{-a/2}^{a/2} {\bf n} \times 
{\bf E}_{mn}({\bf r}') 
e^{-{\rm i} {\bf k} \cdot {\bf r}' } 
\, {\rm d}x'{\rm d}y', \\
& =& A_{mn}
\frac{{\rm i} e^{{\rm i} kr}}{2 \pi r} {\bf k} \times 
\left( 0 , \tilde{E}_{mn}, 0 \right )^T, 
\label{dif}
\end{eqnarray}
where 
\begin{align}
\tilde{E}_{mn} (k_x,k_y) =  \int_{-a/2}^{a/2}  \int_{-a/2}^{a/2}
\cos \left [ \frac{m \pi (x'+a/2)}{a} \right ] \sin \left [ \frac{n \pi (y'+a/2)}{a} \right ] 
e^{-{\rm i}(k_x x'+ k_y y')} \,{\rm d}x' {\rm d}y', \label{nulla}
\end{align}
$k$ is the free-space wavenumber, ${\bf k}=k {\bf r}/|{\bf r}|$, ${\bf n}= (0,0,1)$, 
and $T$ denotes the transpose. Also, a change of variables was applied to shift the origin of the coordinate system to the center of the aperture, and the time dependence has been suppressed for brevity.  
Carrying out the integration in Eq.~(\ref{nulla}) gives, apart from constant phase factors, that 
\begin{eqnarray}
\tilde{E}_{01}(k_x,k_y) &=& \frac{2 \sin(k_x a/2)}{k_x}
\frac{2\pi a}{\pi^2-a^2k_y^2} \cos(k_y a/2), \\
\tilde{E}_{02}(k_x,k_y) &=&  \frac{2 \sin(k_x a/2)}{k_x}
\frac{4\pi a}{4\pi^2-a^2 k_y^2} \sin(k_y a/2), \\
\tilde{E}_{11}(k_x,k_y) &=& \frac{ 2k_x a^2}{\pi^2-a^2 k_x^2} \cos(k_x a/2)
\frac{2\pi a}{\pi^2-a^2 k_y^2} \cos(k_y a/2).
\end{eqnarray}

The radiated intensity $I= {\bf S} \cdot {\bf r}$, with ${\bf S} $ the Poynting vector, is given by the sum of the three modal contributions, i.e., 
\begin{eqnarray}
I (k_x,k_y) \propto (k^2-k_y^2)  \left |A_{01} \tilde{E}_{01} +A_{11} e^{{\rm i}\delta_{1}}\tilde{E}_{11}+  A_{02} e^{{\rm i}\delta_{2}} \tilde{E}_{02} \right |^2,
\label{far}
\end{eqnarray}
where $\delta_{1}$ and $\delta_{2}$ indicate the phase of the ${\rm TE_{11}/TM}_{11}$ and  ${\rm TE}_{02}$ modes  
with respect to the ${\rm TE}_{01}$ mode at the exit plane of the aperture. By changing the phase $\delta_1$ we can steer the radiated intensity along the $x$ 
direction, whereas changing $\delta_2$ steers the radiation in the $y$ direction. This principle is illustrated in Figure~1 for the case of just two modes. 
The left-hand panel (a) shows the individual far-zone fields, $\tilde{E}_{01}$ and $\tilde{E}_{02}$,  that are due to the ${\rm TE}_{01}$ and the ${\rm TE}_{02}$ mode, respectively. 
In panel (b) the radiated intensity generated by the superposed modes is plotted for three values of their relative phase $\delta_2$.  
The maximum steering angle can be increased by increasing the amplitude ratio. An example is shown in panel (c), However, as can be seen, this leads to a broader radiation pattern 
for $\delta_2=\pi/2$ and to side lobes for the other values of $\delta_2$. 
\begin{figure}[h]
\subfigure{\includegraphics[width = 2in]{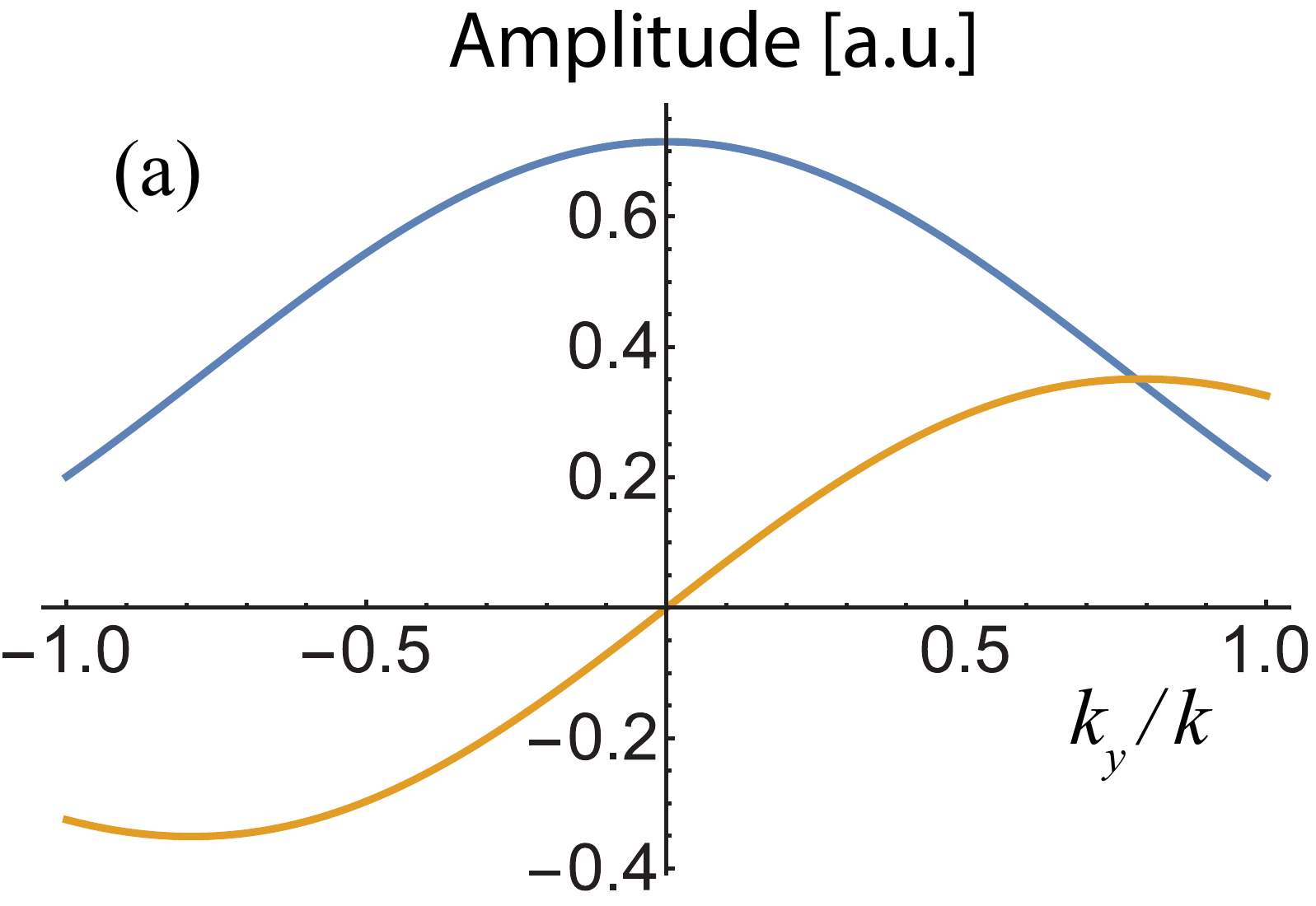}} \quad
\subfigure{\includegraphics[width = 2in]{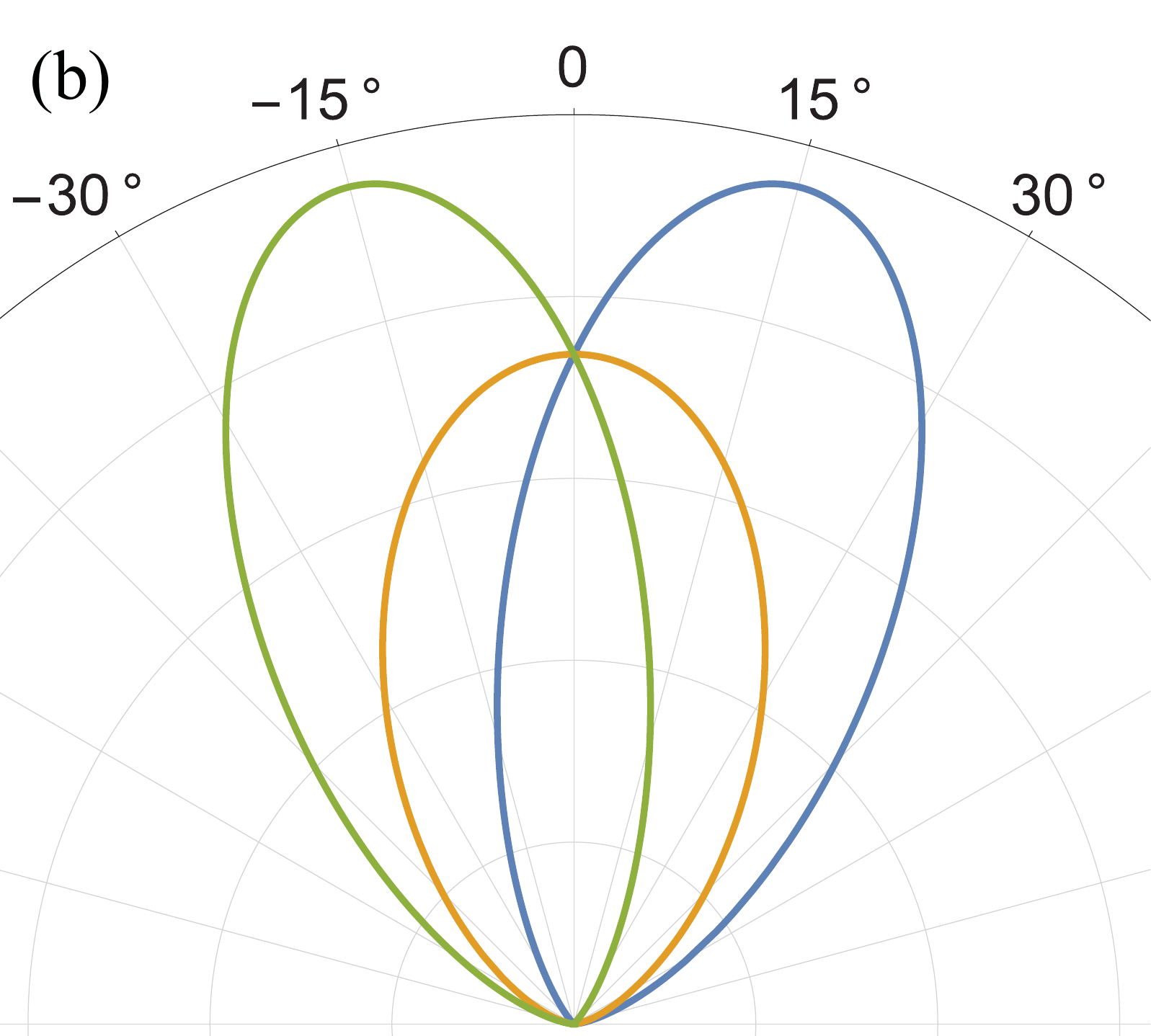}} \quad
\subfigure{\includegraphics[width = 2in]{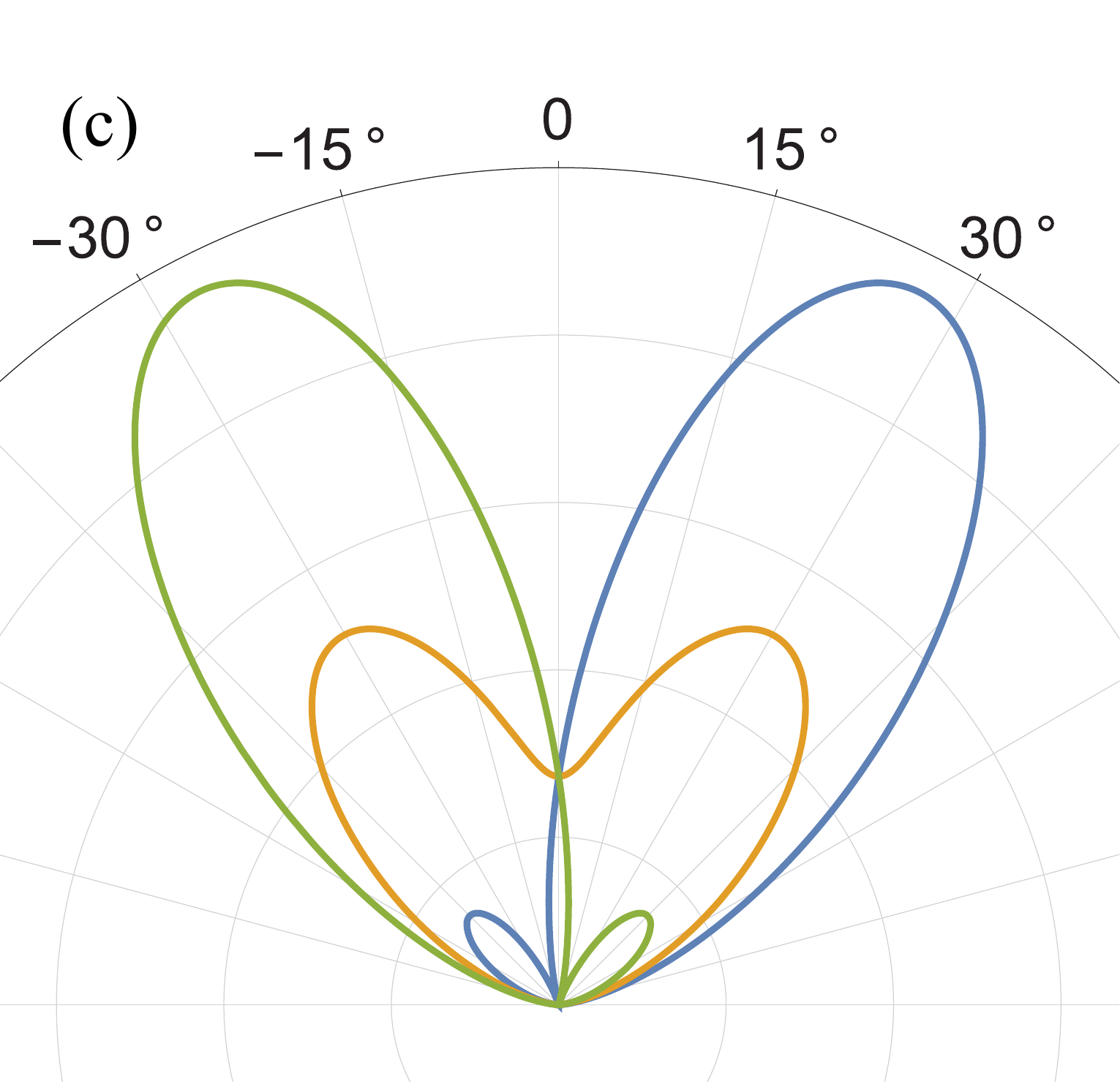}}
\caption{Principle of radiation steering. (a) The far zone field amplitude 
along the $y$ direction due to the  ${\rm TE}_{01}$ mode (blue) and the ${\rm TE}_{02}$ mode (orange).  
(b) Polar plot of the radiated intensity for three values of the relative phase:  $\delta_2= 0$ (blue), $\delta_2= \pi/2$ (orange), and $\delta_2= \pi$ (green). In (a) and (b) the amplitude ratio is set to  $0.6$. In panel (c) this ratio is increased to $2.0$.  
 }
\label{principle}
\end{figure}

\begin{figure}[h]
\includegraphics[scale=0.55]{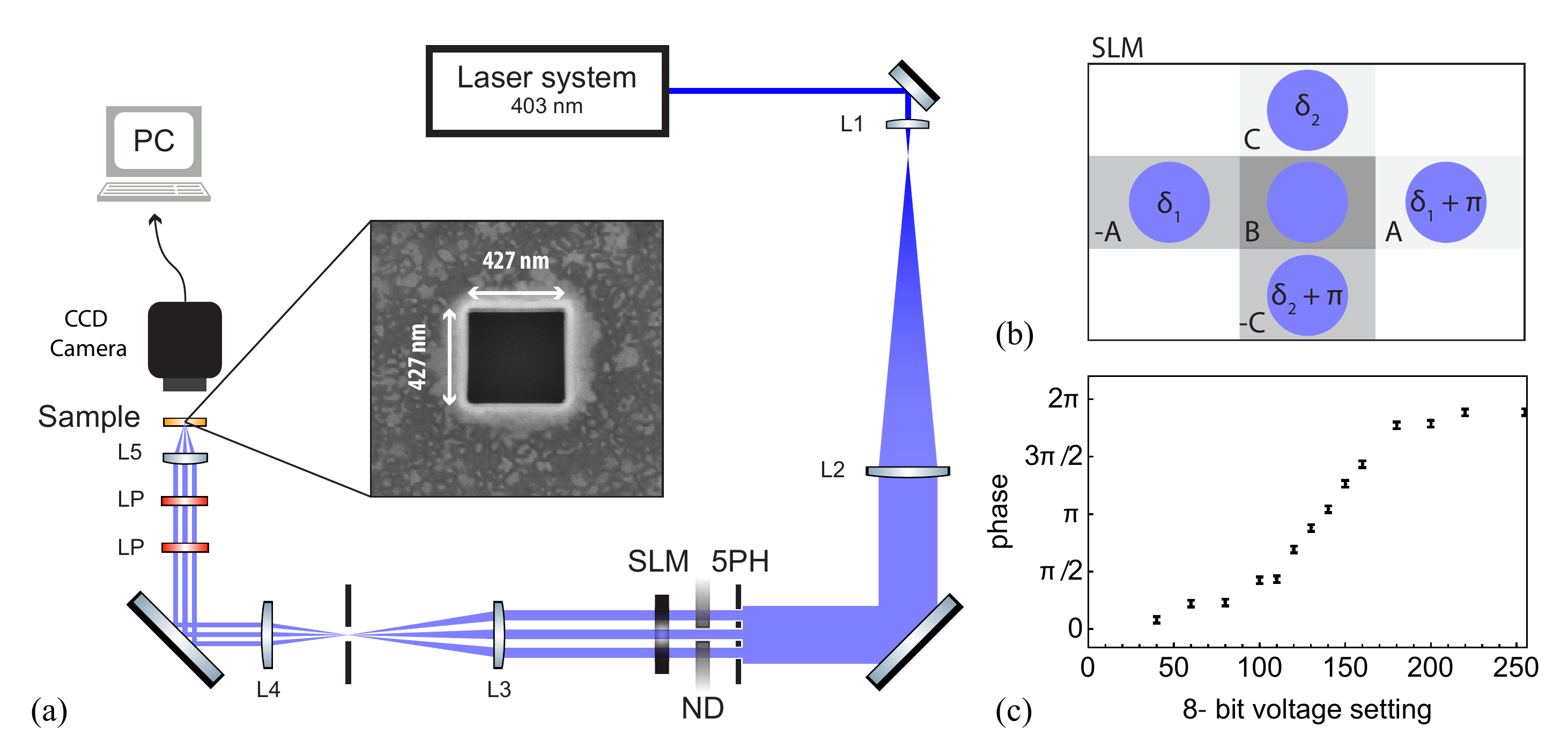}
\caption{Schematic of the setup (a), with an electron microscope image of the gold sample (inset). The optical elements indicated with L1--L5 are lenses, LP are linear polarizers, ND are neutral density filters and 5HP refers to a plate with 5 holes (see text). (b): Layout of the opaque plate with five holes. The fields passing through these holes are labeled 
A, -A, B, C and -C, respectively. 
(c): Calibration curve of the phase change imparted by the SLM as a function of the applied gray level/voltage setting.  }
\label{configuration}
\end{figure}

A sketch of the experimental setup is shown in Figure~2(a). 
A continuous-wave laser was used, operating at $403$~nm, with a bandwidth of $2$~MHz, and an output power of $\sim 300$~mW, of which only a small fraction was used$^{21}$. 
After expansion and beam cleaning the homogeneous, flat wavefront is passed through an opaque 
plate with five identical circular holes, each with a $2$~mm diameter  and a center-to-center distance of $4$~mm (panel~b). 
The plate is put directly in front of a spatial light modulator (SLM) that works in 
transmission (Holoey LC 2012). The SLM is used to imprint a specific phase onto each of the five transmitted beams. 
The induced phase change as a function of the applied gray level (i.e., voltage setting) was determined by observing the transverse 
shift in a two-slit interference pattern. The resulting calibration curve is plotted in panel~(c). 

The inset of panel~(a) shows the sample. It consists of a square aperture $(427\times 427~{\rm nm})$, etched by a focused ion beam, in a $350$~nm thick gold film 
that is mounted on a glass substrate. The aperture size is such that Eq.~(\ref{size}) is satisfied 
for a wavelength of $403$~nm. The film thickness 
ensures that the intensity of the field that  is directly transmitted through the sample is negligable. 

From panel~(b) of Figure~2 it is seen that field B passes through the central hole of the opaque plate 
that covers the SLM. This field is normally incident onto the aperture, and thus has an even symmetry in both the $x$ and $y$ direction. 
According to Table~1, this means that field B only excites 
the ${\rm TE}_{01}$ mode. 
The fields C and -C, which pass through the top and bottom holes, get an imprinted phase of 
$\delta_2$ and $\delta_2+\pi$, respectively. 
The combination of these two fields with opposite phase 
is therefore anti-symmetric in the $y$ direction and symmetric in the $x$ direction. Hence this combined field will only excite 
the  ${\rm TE}_{02}$ mode.   
The phase of the fields A and -A, which travel through the left and right hole,   
is set to $\delta_1$ and $\delta_1 + \pi$, respectively. Because the combination of these two fields 
is anti-symmetric in the $x$ direction and symmetric in the $y$ direction, this will excite just the hybrid ${\rm TE_{11}/TM}_{11}$ mode. 
By changing $\delta_1$ and $\delta_2$  the relative phase of each of the 
three guided modes that are excited in the aperture can be controlled. 
Neutral density filters are placed in front of each outer hole to modulate the intensity of the corresponding beams. A combination of two linear polarizers, before lens 5, is used to modulate the intensity of the central beam and ensure $x$-polarized light. Together this gives us complete control over the superposition of the guided modes, i.e. the total field in the aperture.

\begin{figure}[t]
\includegraphics[scale=0.5]{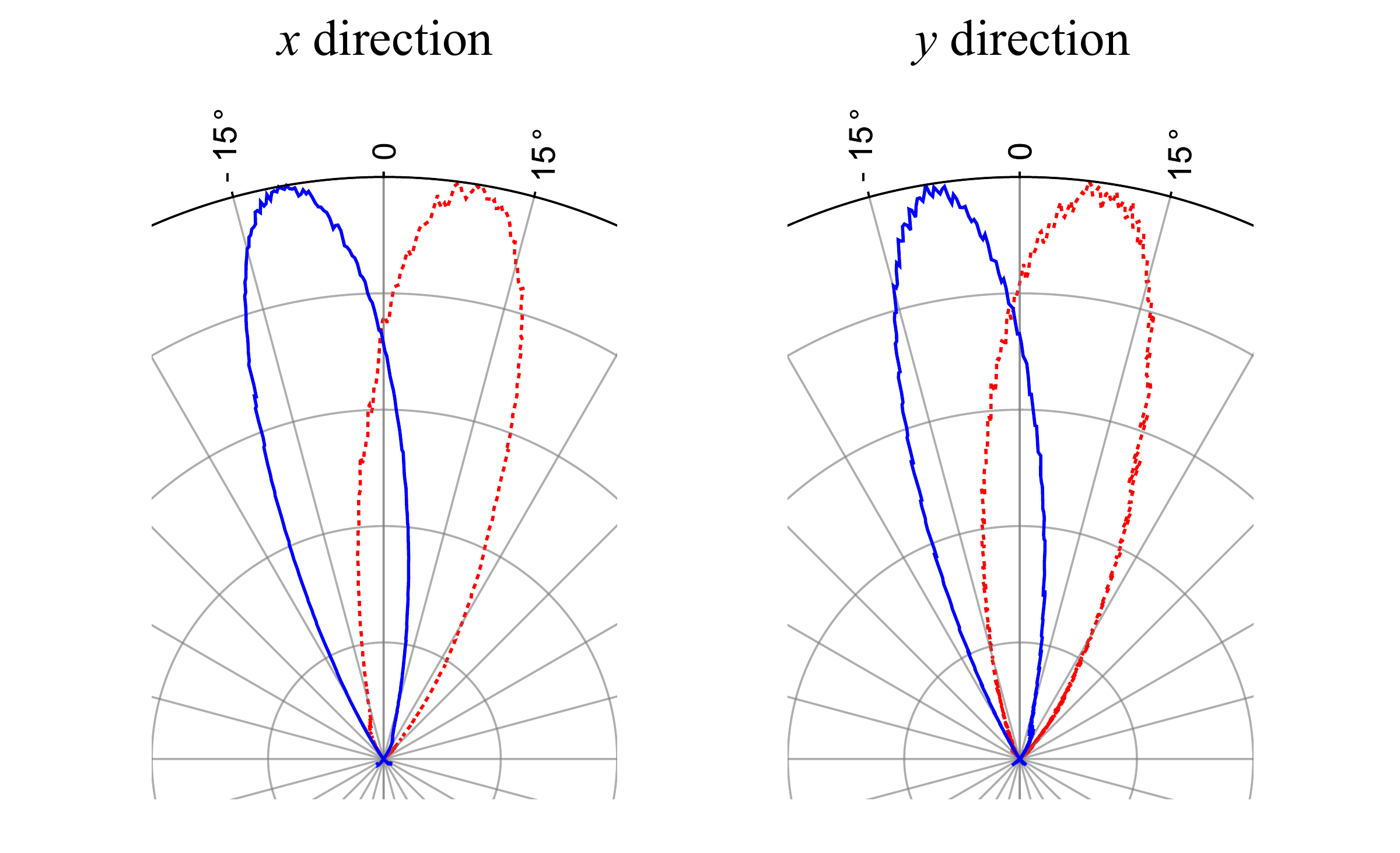}
\caption{(Color online.) Polar plot of the measured radiation pattern for maximum steering in the $x$ and $y$ direction. Zero degrees indicates the forward direction, normal to the metal film. Blue curves: $\delta_1 = 0$ and  $\delta_2 = 0$; 
red curves: $\delta_1 = \pi$ and $\delta_2 = \pi$. 
}
\label{XY}
\end{figure}

\begin{figure}
 \centering
\includegraphics[scale=0.4]{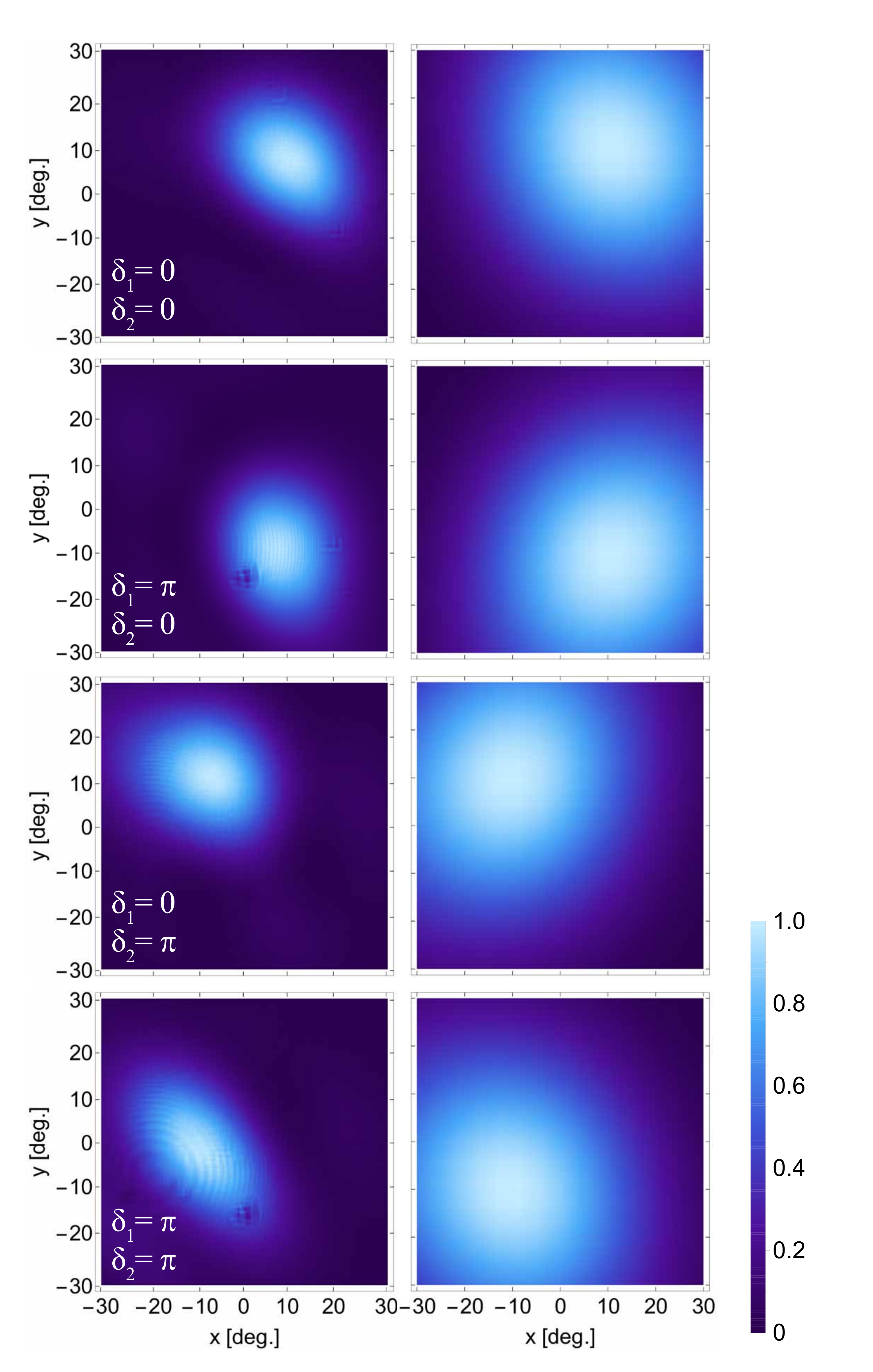}
\caption{(Color online) Observed CCD images (left column) and  simulated images 
(right column) of two-dimensional beam steering. In the simulations $A_{11}=0.6$ and $A_{02}=0.4$. }
\label{2D}
\end{figure}

Our experimental results clearly demonstrate the abilty to steer the radiation. 
Four measured radiation patterns, which are maximally steered in either the $x$ or the $y$ direction, are shown in Figure~3. 
A deflection of $\pm 9.5^\circ$ in both directions was achieved. 
Simultaneous variation of the phases $\delta_{1}$ and $\delta_{2}$ allows us to shift the radiation two-dimensionally in a continuous fashion.  
The setup, built from standard mirror mounts, proved to be very stable, with the radiation pattern undergoing no significant change over a period of hours. This is a significant improvement of 
the stability obtained in$^{19}$.  The efficiency of the setup, i.e., the ratio of the power that is transmitted by the aperture and the the power of the focused field, is largely 
determined by the Fresnel number of L5, the last lens.

Simulated and observed intensities captured with a CCD camera are shown in Figure~4. The left-hand column shows the measured intensity  
patterns for maximum steering angles $-9.5^\circ$ and $9.5^\circ$ in two orthogonal directions. 
The right-hand column shows a simulation, based on Eq.~(\ref{far}), for four different setings of $\delta_1$ and $\delta_2$. The amplitudes were chosen to obtain a reasonable qualitative  agreement between the simulations and the measurements. 
A less than perfect agreement is to be expected, because the simulated results are for the idealized case of a perfectly conducting metal film. This assumption leads to narrower modes and hence 
a broader radiation pattern. Also, the precise value of the mode amplitudes is not known. 
Furthermore, the uncertainty in the angles of observation is approximately $1^\circ$. 

We note that for the particular SLM that was available, a change 
in the phase setting also produced a change in the amplitude and state of polarization.  
Therefore variable neutral density filters were used 
to ensure that the mode amplitudes remained approximately equal after $\delta_1$ and $\delta_2$ had been  varied. The use of a more advanced SLM device$^{22,23}$ 
would make this amplitude adjustment and correction for polarization changes unnecessary. 

In our earlier work$^{19}$ we obtained one-dimensional steering from a nano-slit. 
The current setup is much simpler, involves only one optical path and is 
significantly more stable. A piezo element is no longer required, and 
the use of a thicker film has solved the issue of direct transmission leaking through the sample. 
And, of course, dynamic, two-dimensional steering is now obtained.

We emphasize that our experiment provides a proof of principle of active, two-dimensional beam steering. 
The effect that we observe is a linear optical phenomenon, meaning that a high power laser, as was used in our setup, is not necessary. The phenomenon of beam steering is scalable to the low power levels that may be produced and handled ``on-chip" in semiconductor devices. In our experiment a laser with a very long coherence length (2 MHz optical bandwidth) was used. However, because our setup essentially uses a single optical path, a shorter 
coherence length would also suffice. 
The $403$~nm wavelength was chosen to obtain the full $2\pi$ phase range of the 
specific SLM that was used.   Our technique can be applied at different wavelengths, provided that 
the inequalities~(\ref{size}), that relate the aperture size $a$ to the wavelength $\lambda$, are satisfied and the SLM provides a full $2\pi$ phase range. 
For example, for the choice of a telecom wavelength $\lambda =1500$~nm, the required value of the aperture size would be $1500~{\rm nm} \le a \le 1677~{\rm nm}$. 
It is worth noting that the maximum steering angle can be tuned by changing the relative amplitudes of the modes, although larger steering angles are accompanied by a broadening of the 
radiation pattern and the onset of side lobes as was shown in Figure~1.

In conclusion, we have demonstrated dynamic control of the direction of radiation that emanates from a narrow square aperture in a metal film. 
This was accomplished by selective excitation of the three guided modes that such an aperture allows. 
Our method  uses only an SLM, and does not involve any mechanical adjustment of optical elements.   
A simple wave-guiding model provides physical understanding and, even though it assumes perfect conductivity, 
gives good qualitative agreement with the experimental results.  
Unlike previously reported static configurations$^{15-17}$, our dynamic setup can be used 
as an optical switch in photonic circuitry in which light is sent to different ports, or in optical biosensors in which 
samples need to be scanned. 

\vspace*{0.50cm}

The authors declare no competing financial interest. 

\subsection*{Acknowledgements}
The authors wish to thank Andries Lof of AMOLF NanoCenter Amsterdam for fabrication of the sample. 
T.D.V. acknowledges support from the Air Force Office of Scientific Research under award number FA9550-16-1-0119.



\end{document}